\documentclass[aps,twocolumn,showpacs]{revtex4}
\usepackage{epsfig}
\usepackage{amsbsy}
\usepackage{amsmath}
\usepackage{amsfonts}
\usepackage{graphicx}
\usepackage{latexsym}

\begin{document}

\title{Entanglement in spin-1/2 dimerized Heisenberg systems}
\author{Zhe Sun, XiaoGuang Wang, AnZi Hu, and You-Quan Li}
\affiliation{Zhejiang Institute of Modern Physics, Department of
Physics, Zhejiang University, HangZhou 310027, China}
\date{\today}
\begin{abstract}
We study entanglement in dimerized Heisenberg systems. In
particular, we give exact results of ground-state pairwise
entanglement for the four-qubit model by identifying a $Z_2$
symmetry. Although the entanglements cannot identify the critical
point of the system, the mean entanglement of nearest-neighbor
qubits really does, namely, it reaches a maximum at the critical
point.
\end{abstract}
 \maketitle

The study of entanglement properties of many-body condensed matter
systems has received much
attention~\cite{M_Nielsen}-\cite{QPT_GVidal}. It covers both the
rapid developing area of quantum information and the area of
condensed matter. Various quantum models have been considered in
the literature and some of them are exactly solvable.
Specifically, it was well-known that the anisotropic Heisenberg
model can be solved formally by Bethe's Ansatz
method~\cite{Bethe,Yang} for arbitrary number of qubits $N$,
however, we have to solve a set of transcendental equations.

It is interesting to see that entanglement in a system with a few
sites displays general features of entanglement with more sites.
In the anisotropic Heisenberg model with a large number of qubits,
the pairwise entanglement shows a maximum in the isotropic
point~\cite{Gu}. This feature was already shown in a small system
with four or five qubits~\cite{Wang04pla1}. So, the study of small
systems is meaningful in the study of entanglement as they can
reflect general features of larger or macroscopic systems.

Dimerized systems play an important role in condensed matter
theory,  and entanglement in dimerized systems has been
considered~\cite{Chen,Vedraltwo}. Here, we aim at obtaining some
analytical results of entanglement in a small system. Chen et
al~\cite{Chen} found that the concurrence~\cite{Conc}, quantifying
entanglement of a pair of qubits cannot enable us to identify the
critical point $J_2/J_1=1$. However, we will show that although
the single concurrence cannot be used to identify the critical
point, the mean concurrence really does.

For $N\le 7$, the isotropic Heisenberg Hamiltonian can be
analytically solved~\cite{Kouzoudis,Schnack} due to the
translational invariant symmetry which can be used to reduce the
Hamiltonian matrix to smaller submatrices by a factor of
$N$~\cite{Lin}. However, for our dimerized systems, the
translational invariant symmetry breaks. We identify a $Z_2$
symmetry in the dimerized model, and using this symmetry we can
further reduce Hamiltonian matrices by a factor of two. We will
first consider the simplest representative four-qubit dimerized
system, and by using the $Z_2$ symmetry, all eigenvalues can be
obtained.

{\em Eigenvalue problem.} The Heisenberg Hamiltonian for the
dimerized chain with even number of qubits $N$ reads
\begin{align}
H=&\sum_{i=1}^{N/2}\left[J_1(2{\bf S}_{2i-1}\cdot {\bf
S}_{2i}+\frac{1}2)+J_2(2{\bf S}_{2i}\cdot
{\bf S}_{2i+1}+\frac{1}2)\right]\nonumber\\
=&\sum_{i=1}^{N/2}(J_1 {\cal S}_{2i-1,2i}+J_2 {\cal S}_{2i,2i+1}).
 \label{H}
\end{align}
where ${\bf S}_i$ is the spin-half operator for qubit $i$ ,
${\cal S}_{j,j+1}=\frac 12\left( 1+\vec{\sigma}_i\cdot \vec{\sigma}%
_{i+1}\right) $ is the swap operator between qubit $i$ and $j$, $\vec{\sigma}%
_i=(\sigma _{ix},\sigma _{iy},\sigma _{iz})$ is the vector of
Pauli matrices, and $J_1$ and $J_2$ are the exchange constants. We
have assumed the periodic boundary condition, i.e., $N+1\equiv 1$.
In the following discussions, we also assume $J_1,J_2\ge 0$
(antiferromagnetic case).

For the four-qubit case, the dimerized Hamiltonian simplifies to
\begin{equation}
H=J_1({\cal S}_{12}+{\cal S}_{34})+J_2 ({\cal S}_{23}+{\cal
S}_{41}). \label{HH}
\end{equation}
Although we have imposed the periodic boundary condition, the
Hamiltonian is no longer translational invariant except the case
of $J_1=J_2$. As we stated above, the translational invariant
symmetry can be used to reduce the Hamiltonian matrix to smaller
submatrices by a factor of $N$~\cite{Lin}. However, for our case,
we cannot have this reduction.

To exactly solve the eigenvalue problem of the Heisenberg model,
we first note that $[H,J_z]=0$, the whole 16-dimensional Hilbert
space can be divided into invariant subspaces spanned by vectors
with a fixed number of reversed spins. Then, the largest subspace
is 6-dimensional with 2 reversed spins. Here,
$J_z=\sum_{i=1}^4\sigma_{iz}/2$. Due to the $Z_2$ symmetry
$[H,\Sigma_x]=0$, it is sufficient to solve the eigenvalue
problems in the subspaces with $r$ reversed spins, where
$r\in\{0,1,2\}$ and
\begin{equation}
\Sigma_x=\sigma_x\otimes\sigma_x\otimes\sigma_x\otimes\sigma_x.
\end{equation}

The subspace with $r=0$ only contains one vector $|0000\rangle$,
which is the eigenvector with eigenvalue
\begin{equation}
E_0=2(J_1+J_2).\label{e1}
\end{equation}
For subspace with $2\ge r>0$, the smallest submatrix is $4\times
4$. We need further reduce submatrices to smaller ones. Although
we do not have a translational invariance, from the four-qubit
Hamiltonian (\ref{HH}), we can identify another symmetry $Z_2$
symmetry, given by
\begin{equation}
[H,{\cal S}_{12,34}]=[H,{\cal S}_{13}{\cal S}_{24}]=0.
\end{equation}
The operator exchange the state of qubits 1 and 2 with the state
of qubits 3 and 4, namely,
\begin{equation}
{\cal S}_{12,34}|m_1,m_2,m_3,m_4\rangle=|m_3,m_4,m_1,m_2\rangle.
\end{equation}
The eigenvalue of the operator ${\cal S}_{12,34}$ on all energy
eigenstates is either 1 or -1.

The subspace with $r=1$ is spanned by four basis vectors
$\{|1000\rangle,|0100\rangle,|0010\rangle,|0001\rangle \}$. Taking
into account the $Z_2$ symmetry, we choose the following basis
\begin{align}
|\psi_{1\pm}\rangle=\frac{1}{\sqrt{2}}(|1000\rangle\pm|0010\rangle),\nonumber\\
|\phi_{1\pm}\rangle=\frac{1}{\sqrt{2}}(|0100\rangle\pm|0001\rangle).
\end{align}
Obviously, they are all eigenstates of operator ${\cal
S}_{12,34}$, and basis $\{|\psi_+\rangle,|\phi_+\rangle\}$
($\{|\psi_-\rangle,|\phi_-\rangle\}$) spans an invariant
two-dimensional subspace of Hamiltonian $H$. Let the Hamiltonian
act on the basis, then we obtain
\begin{align}
H|\psi_{1+}\rangle=&(J_1+J_2)(|\psi_{1+}\rangle+\phi_{1+}\rangle), \nonumber\\
H|\phi_{1+}\rangle=&(J_1+J_2)(|\psi_{1+}\rangle+\phi_{1+}\rangle), \nonumber\\
H|\psi_{1-}\rangle=&(J_1+J_2)|\psi_{1-}\rangle+(J_1-J_2)|\phi_{1-}\rangle, \nonumber\\
H|\phi_{1-}\rangle=&(J_1+J_2)|\phi_{1-}\rangle+(J_1-J_2)|\psi_{1-}\rangle.
\end{align}
From the above equation, the eigenvalues of Hamiltonian $H$ is
obtained as
\begin{equation}
E_{1,1}=2(J_1+J_2), E_{1,2}=0, E_{1,3}=2J_1, E_{1,4}=2J_2,
\end{equation}
where the first subscript denotes the number of reversed spins.

For the case of $r=2$, due to the existence of the $Z_2$ symmetry,
we choose the following basis for the six-dimensional subspace
\begin{align}
|\psi_{2\pm}\rangle=\frac{1}{\sqrt{2}}(|1100\rangle\pm|0011\rangle),\nonumber\\
|\phi_{2\pm}\rangle=\frac{1}{\sqrt{2}}(|1001\rangle\pm|0110\rangle),\nonumber\\
|\varphi_{2\pm}\rangle=\frac{1}{\sqrt{2}}(|0101\rangle\pm|1010\rangle).
\end{align}
Then, we can reduce the $6\times 6$ Hamiltonian matrix to a
block-diagonal form with two $3\times 3$ matrices. After the
action of the Hamiltonian on the above basis, we obtain
\begin{align}
H|\psi_{2-}\rangle=&2J_1|\psi_{2-}\rangle,\nonumber\\
H|\phi_{2-}\rangle=&2J_2|\phi_{2-}\rangle,\nonumber\\
H|\varphi_{2-}\rangle=&0,\nonumber\\
H|\psi_{2+}\rangle=&2J_1|\psi_{2+}\rangle+2J_2|\varphi_{2+}\rangle,\nonumber\\
H|\phi_{2+}\rangle=&2J_2|\phi_{2+}\rangle+2J_1|\varphi_{2+}\rangle,\nonumber\\
H|\varphi_{2+}\rangle=&2J_2|\psi_{2+}\rangle+2J_1|\phi_{2+}\rangle.
\label{r2}
\end{align}
One $3\times 3$ block is of the diagonal form, and the eigenvalues
read
\begin{equation}
E_{2,1}=2J_1,\;E_{2,2}=2J_2,\;E_{2,3}=0.
\end{equation}
Another block is written as
\begin{equation}
H=\left(
\begin{array}{lll}
2J_1&0&2J_2\\
0&2J_2&2J_1\\
2J_2&2J_1&0
\end{array}
\right). \label{HHH}
\end{equation}
Although this is a $3\times 3$ matrix, we can further reduce it to
$2\times 2$ matrix since, from the last three equations of
Eq.~(\ref{r2}), one eigenvalue is found to be
$E_{2,4}=2(J_1+J_2)$. With the help is this eigenvalue, the
characteristic polynomial of the Hamiltonian matrix can be brought
to a quadratic form by dividing it with $E-E_{2,4}$. Then, we
obtain
\begin{align}
E_{2,5}=&2\sqrt{J_1^2+J_2^2-J_1J_2},\nonumber\\
E_{\text{GS}}=&E_{2,6}=-2\sqrt{J_1^2+J_2^2-J_1J_2}.
\end{align}

Thus, all eigenvalues are analytically obtained for the four-spin
dimerized Heisenberg model. We see that ground state is
non-degenerate and the energy is given by $E_{2,6}$. Although the
eigenstates can be easily obtained, they are not given explicitly
here as the knowledge of eigenvalues is sufficient for discussions
of entanglement properties as we will see below.

{\em Entanglement in the four-qubit model.} We first study the
ground-state entanglement of qubits 1 and 2. Due to the SU(2)
symmetry in our Hamiltonian, the concurrence quantifying the
entanglement of two qubits is given by~\cite{Wang04}
\begin{equation}
C_{i,i+1}=-2{\langle {\bf S}_i\cdot {\bf
S}_{i+1}\rangle}-1/2=-\langle {\cal S}_{i,i+1}\rangle,
\label{cccc}
\end{equation}
where we have ignored the max function in the usual definition of
the concurrence, and thus the negative concurrence implies no
entanglement. We see that the entanglement is determined by the
expectation value of the SWAP operator. There are two distinctive
concurrences $C_{12}$ and $C_{23}$, corresponding to two spins
coupled by bond $J_1$ and $J_2$.

The expectation value $\langle {\cal S}_{12}\rangle$ can be
calculated via Feynman-Hellman theorem. By applying the theorem to
the ground state, we obtain
\begin{equation}
\langle {\cal S}_{12}\rangle+\langle {\cal
S}_{34}\rangle=\frac{\partial E_{\text{GS}}}{\partial
J_1}=\frac{J_2-2J_1}{\sqrt{J_1^2+J_2^2-J_1J_2}}. \label{sss}
\end{equation}
Due to the $Z_2$ symmetry $[H,{\cal S}_{12,34}]$, we have $
\langle{\cal S}_{12}\rangle=\langle{\cal S}_{34}\rangle$. Then,
from the above equation, the expectation value $\langle{\cal
S}_{12}\rangle$ is obtained. Substituting it to Eq.~(\ref{cccc})
leads to
\begin{equation}
C_{12}=\frac{2J_1-J_2}{2\sqrt{J_1^2+J_2^2-J_1J_2}}.\label{a1}
\end{equation}
In a similar way, we can obtain the concurrence $C_{23}=C_{14}$ as
\begin{equation}\label{a2}
C_{23}=\frac{2J_2-J_1}{2\sqrt{J_1^2+J_2^2-J_1J_2}}.
\end{equation}
Thus, the analytical expressions of two types of concurrence are
obtained, and the concurrence $C_{12}$ can be transformed to
$C_{23}$ by exchanging $J_1$ and $J_2$.

From Eqs.~(\ref{a1}) and (\ref{a2}), we read that
\begin{equation}
{C}_{12}\left\{
\begin{array}{ll}
=1  & \text{when} ~~J_2=0, \\
=1/2& \text{when} ~~J_2=J_1, \\
\le 0  & \text{when} ~~J_2\ge J_{\text{2th}}^{(12)}=2J_1,
\end{array}
\right.   \label{c12}
\end{equation}
and
\begin{equation}
{C}_{23}\left\{
\begin{array}{ll}
\le 0  & \text{when} ~~J_2\le J_{\text{2th}}^{(23)}=J_1/2, \\
=1/2& \text{when} ~~J_2=J_1, \\
=1  & \text{when} ~~J_2=\infty,
\end{array}
\right.   \label{c23}
\end{equation}
In the case of $J_2=0$, the ground state is just the uncoupled two
singlets, and thus $C_{12}=1$. When $J_2=J_1$, the dimerized
Hamiltonian is reduced to the isotropic one, and the concurrence
$C_{12}=C_{23}=1/2$~\cite{WangPaolo}. When
$J_2>J_{\text{2th}}^{(23)}$, the entanglement between qubits 2 and
3 builds up, and when $J_2\ge J_{\text{2th}}^{(12)}=2J_1$, the
entanglement between qubits 1 and 2 vanishes.

\begin{figure}
\includegraphics[width=0.45\textwidth]{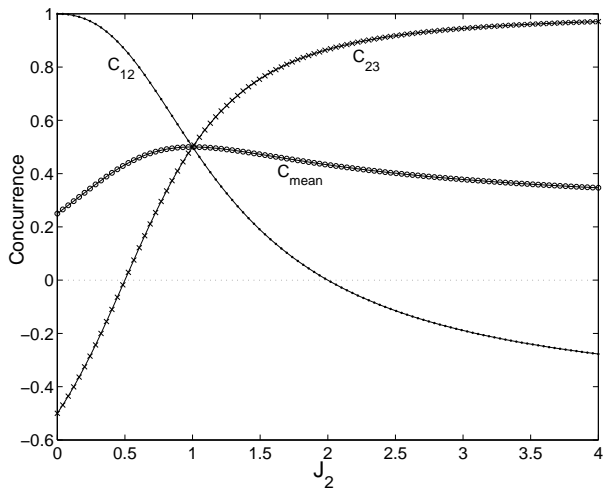}
\caption{The concurrences versus $J_2$ in the four-qubit dimerized
model ($J_1=1$).}
\end{figure}
It was argued that the pairwise entanglements quantified by the
concurrence cannot identify the critical point ($J_2=J_1$) of the
dimerized system~\cite{Chen}, however here we use the mean
entanglement of nearest-neighbor qubits as the system we are
considering is not a uniform system, namely, it has two distinct
exchange constants. There are four pairs of nearest-neighbor
qubits, and thus
\begin{equation}
C_\text{mean}=\frac{C_{12}+C_{23}+C_{34}+C_{41}}4=\frac{C_{12}+C_{23}}2.\label{meann}
\end{equation}
Substituting Eqs.~(\ref{a1}) and (\ref{a2}) to the above equation
leads to
\begin{equation}\label{mean}
C_\text{mean}=\frac{J_1+J_2}{4\sqrt{J_1^2+J_2^2-J_1J_2}},
\end{equation}
from which we read
\begin{equation} {C}_{\text{mean}}=\left\{
\begin{array}{ll}
1/4  & \text{when} ~~J_2=0, \\
1/2& \text{when} ~~J_2=J_1, \\
1/4  & \text{when} ~~J_2=\infty.
\end{array}
\right.   \label{c23}
\end{equation}
In Fig.~1, we plot the concurrences and the mean concurrence
versus $J_2$. We see that the mean entanglement takes its maximum
at the critical point $J_2$=$J_1$. From Eq.~(\ref{mean}), we can
also check that the maximal point is at the critical point. In the
following, we will numerically show that this fact still holds for
dimerized chains with more qubits.

\begin{figure}
\includegraphics[width=0.45\textwidth]{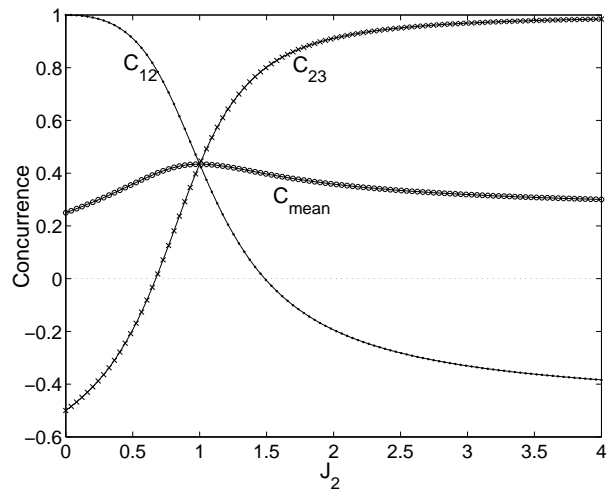}
\caption{The concurrences versus $J_2$ in the six-qubit dimerized
model ($J_1=1$).}
\end{figure}

\begin{figure}
\includegraphics[width=0.45\textwidth]{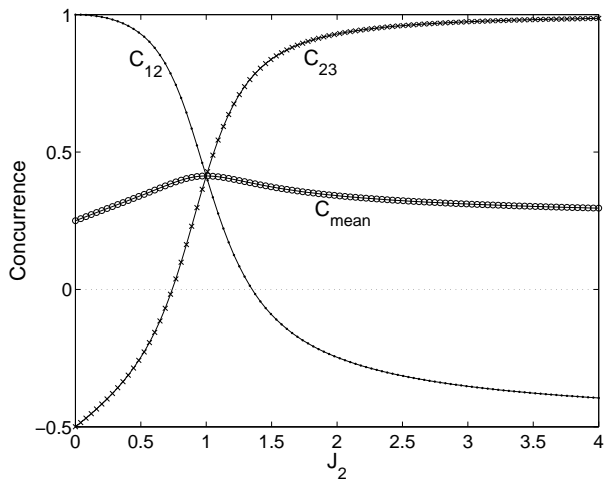}
\caption{The concurrences versus $J_2$ in the eight-qubit
dimerized model ($J_1=1$).}
\end{figure}

{\em Numerical results.} Now we consider more general cases of
even-number qubits. For even $N>4$ , the analytical results of
entanglement are hard to obtain, and here, we use exact
diagonalization method to numerically calculate the entanglement.
For even-number case, the mean entanglement is still given by
Eq.~(\ref{meann}).

In Figs.~2 and 3, we plot the concurrences versus $J_2$ for $N=6$
and $N=8$, respectively. We observe that the entanglement
properties are similar to those in the four-qubit model, namely,
they are qualitatively the same. The mean entanglement reaches its
maximum at the critical point. From Figs.~1-3, we see that the
threshold value $J_{2\text{th}}^{(12)}$ decreases, while
$J_{2\text{th}}^{(23)}$ increases when the number of qubits
increases.

Finally, we give a relation between the two types of entanglement
with the ground-state energy $E_\text{GS}$. From Eq.~(\ref{H}), we
immediately have the relation between the ground-state energy and
the correlators $\langle{\cal S}_{12}\rangle$ and $\langle{\cal
S}_{23}\rangle$,
\begin{equation}\label{r}
E_\text{GS}/N=\frac{1}2 (J_1\langle{\cal
S}_{12}\rangle+J_2\langle{\cal S}_{23}\rangle).
\end{equation}
Then, from Eq.~(\ref{cccc}), we obtain
\begin{equation}
E_{\text GS}/N=-\frac{1}2(J_1C_{12}+J_2C_{23}).
\end{equation}
The above equation gives the relation between ground-state energy
and the two typical concurrences. If we know the entanglement of
qubits 1 and 2 and the ground-state energy, we can know the
entanglement of qubits 2 and 3 from the relation.

{\em Conclusion.} We have studied entanglement in spin-1/2
dimerized Heisenberg systems. As a representative system, the
four-qubit model was studied in detail. By identifying the $Z_2$
symmetry, we have solved the eigenvalue problem completely. From
the ground-state energy, the analytical results of the two types
of pairwise entanglement were obtained. We have located the
threshold value of $J_2$ after which the entanglement between
qubits 1 and 2 vanishes and the threshold value before which there
exists no entanglement between qubits 2 and 3.

In order to identify the critical point of the dimerized system,
we have proposed the mean entanglement of nearest-neighbor qubits
as an efficient indicator. The mean entanglement displays a
maximum at the critical point. We have also considered the case of
more qubits, and the numerical results show that the entanglement
properties are similar to those in the four-qubit model. Dimerized
Heisenberg systems play an important role in condensed matter
physics. It is interesting to study entanglement in other
dimerized systems, which is under consideration.

\acknowledgements We thanks for the helpful discussions with Yan
Chen.

\end{document}